\documentclass[12pt]{article}
\usepackage{amsfonts}

\makeatletter
\newcount\@hour\newcount\@minute\chardef\@x10\chardef\@xv60
\def\tcitime{
\def\@time{%
  \@minute\time\@hour\@minute\divide\@hour\@xv
  \ifnum\@hour<\@x 0\fi\the\@hour:%
  \multiply\@hour\@xv\advance\@minute-\@hour
  \ifnum\@minute<\@x 0\fi\the\@minute
  }}%
\def\QCTOpt[#1]#2{%
  \def\QCTOptB{#1}
  \def\QCTOptA{#2}
}
\def\QCTNOpt#1{%
  \def\QCTOptA{#1}
  \let\QCTOptB\empty
}
\def\Qct{%
  \@ifnextchar[{%
    \QCTOpt}{\QCTNOpt}
}
\def\QCBOpt[#1]#2{%
  \def\QCBOptB{#1}
  \def\QCBOptA{#2}
}
\def\QCBNOpt#1{%
  \def\QCBOptA{#1}
  \let\QCBOptB\empty
}
\def\Qcb{%
  \@ifnextchar[{%
    \QCBOpt}{\QCBNOpt}
}
\def\PrepCapArgs{%
  \ifx\QCBOptA\empty
    \ifx\QCTOptA\empty
      {}%
    \else
      \ifx\QCTOptB\empty
        {\QCTOptA}%
      \else
        [\QCTOptB]{\QCTOptA}%
      \fi
    \fi
  \else
    \ifx\QCBOptA\empty
      {}%
    \else
      \ifx\QCBOptB\empty
        {\QCBOptA}%
      \else
        [\QCBOptB]{\QCBOptA}%
      \fi
    \fi
  \fi
}
\newcount\GRAPHICSTYPE
\GRAPHICSTYPE=\z@
\def\GRAPHICSPS#1{%
 \ifcase\GRAPHICSTYPE
   \special{ps: #1}%
 \or
   \special{language "PS", include "#1"}%
 \fi
}%
\def\graffile#1#2#3#4{%
    \leavevmode
    \raise -#4 \BOXTHEFRAME{%
        \hbox to #2{\raise #3\hbox{\null #1}}}%
}%
\def\draftbox#1#2#3#4{%
 \leavevmode\raise -#4 \hbox{%
  \frame{\rlap{\protect\tiny #1}\hbox to #2%
   {\vrule height#3 width\z@ depth\z@\hfil}%
  }%
 }%
}%
\newcount\draft
\draft=\z@

\newif\ifwasdraft
\wasdraftfalse

\def\GRAPHIC#1#2#3#4#5{%
 \ifnum\draft=\@ne\draftbox{#2}{#3}{#4}{#5}%
  \else\graffile{#1}{#3}{#4}{#5}%
  \fi
 }%
\def\addtoLaTeXparams#1{%
    \edef\LaTeXparams{\LaTeXparams #1}}%
\newif\ifBoxFrame \BoxFramefalse
\newif\ifOverFrame \OverFramefalse
\newif\ifUnderFrame \UnderFramefalse

\def\BOXTHEFRAME#1{%
   \hbox{%
      \ifBoxFrame
         \frame{#1}%
      \else
         {#1}%
      \fi
   }%
}

\def\doFRAMEparams#1{\BoxFramefalse\OverFramefalse\UnderFramefalse\readFRAMEparams#1\end}%
\def\readFRAMEparams#1{%
 \ifx#1\end%
  \let\next=\relax
  \else
  \ifx#1i\dispkind=\z@\fi
  \ifx#1d\dispkind=\@ne\fi
  \ifx#1f\dispkind=\tw@\fi
  \ifx#1t\addtoLaTeXparams{t}\fi
  \ifx#1b\addtoLaTeXparams{b}\fi
  \ifx#1p\addtoLaTeXparams{p}\fi
  \ifx#1h\addtoLaTeXparams{h}\fi
  \ifx#1X\BoxFrametrue\fi
  \ifx#1O\OverFrametrue\fi
  \ifx#1U\UnderFrametrue\fi
  \ifx#1w
    \ifnum\draft=1\wasdrafttrue\else\wasdraftfalse\fi
    \draft=\@ne
  \fi
  \let\next=\readFRAMEparams
  \fi
 \next
 }%
\def\IFRAME#1#2#3#4#5#6{%
      \bgroup
      \let\QCTOptA\empty
      \let\QCTOptB\empty
      \let\QCBOptA\empty
      \let\QCBOptB\empty
      #6%
      \parindent=0pt%
      \leftskip=0pt
      \rightskip=0pt
      \setbox0 = \hbox{\QCBOptA}%
      \@tempdima = #1\relax
      \ifOverFrame
          \typeout{This is not implemented yet}%
          \show\HELP
      \else
         \ifdim\wd0>\@tempdima
            \advance\@tempdima by \@tempdima
            \ifdim\wd0 >\@tempdima
               \textwidth=\@tempdima
               \setbox1 =\vbox{%
                  \noindent\hbox to \@tempdima{\hfill\GRAPHIC{#5}{#4}{#1}{#2}{#3}\hfill}\\%
                  \noindent\hbox to \@tempdima{\parbox[b]{\@tempdima}{\QCBOptA}}%
               }%
               \wd1=\@tempdima
            \else
               \textwidth=\wd0
               \setbox1 =\vbox{%
                 \noindent\hbox to \wd0{\hfill\GRAPHIC{#5}{#4}{#1}{#2}{#3}\hfill}\\%
                 \noindent\hbox{\QCBOptA}%
               }%
               \wd1=\wd0
            \fi
         \else
            \ifdim\wd0>0pt
              \hsize=\@tempdima
              \setbox1 =\vbox{%
                \unskip\GRAPHIC{#5}{#4}{#1}{#2}{0pt}%
                \break
                \unskip\hbox to \@tempdima{\hfill \QCBOptA\hfill}%
              }%
              \wd1=\@tempdima
           \else
              \hsize=\@tempdima
              \setbox1 =\vbox{%
                \unskip\GRAPHIC{#5}{#4}{#1}{#2}{0pt}%
              }%
              \wd1=\@tempdima
           \fi
         \fi
         \@tempdimb=\ht1
         \advance\@tempdimb by \dp1
         \advance\@tempdimb by -#2%
         \advance\@tempdimb by #3%
         \leavevmode
         \raise -\@tempdimb \hbox{\box1}%
      \fi
      \egroup%
}%
\def\DFRAME#1#2#3#4#5{%
 \begin{center}
     \let\QCTOptA\empty
     \let\QCTOptB\empty
     \let\QCBOptA\empty
     \let\QCBOptB\empty
     \ifOverFrame 
        #5\QCTOptA\par
     \fi
     \GRAPHIC{#4}{#3}{#1}{#2}{\z@}
     \ifUnderFrame 
        \par #5\QCBOptA
     \fi
 \end{center}%
 }%
\def\FFRAME#1#2#3#4#5#6#7{%
 \begin{figure}[#1]%
  \let\QCTOptA\empty
  \let\QCTOptB\empty
  \let\QCBOptA\empty
  \let\QCBOptB\empty
  \ifOverFrame
    #4
    \ifx\QCTOptA\empty
    \else
      \ifx\QCTOptB\empty
        \caption{\QCTOptA}%
      \else
        \caption[\QCTOptB]{\QCTOptA}%
      \fi
    \fi
    \ifUnderFrame\else
      \label{#5}%
    \fi
  \else
    \UnderFrametrue%
  \fi
  \begin{center}\GRAPHIC{#7}{#6}{#2}{#3}{\z@}\end{center}%
  \ifUnderFrame
    #4
    \ifx\QCBOptA\empty
      \caption{}%
    \else
      \ifx\QCBOptB\empty
        \caption{\QCBOptA}%
      \else
        \caption[\QCBOptB]{\QCBOptA}%
      \fi
    \fi
    \label{#5}%
  \fi
  \end{figure}%
 }%
\newcount\dispkind%
\def\FRAME#1#2#3#4#5#6#7#8{%
 \ifnum\draft=\@ne
   \wasdrafttrue
 \else
   \wasdraftfalse%
 \fi
 \def\LaTeXparams{}%
 \dispkind=\z@
 \def\LaTeXparams{}%
 \doFRAMEparams{#1}%
 \ifnum\dispkind=\z@\IFRAME{#2}{#3}{#4}{#7}{#8}{#5}\else
  \ifnum\dispkind=\@ne\DFRAME{#2}{#3}{#7}{#8}{#5}\else
   \ifnum\dispkind=\tw@
    \edef\@tempa{\noexpand\FFRAME{\LaTeXparams}}%
    \@tempa{#2}{#3}{#5}{#6}{#7}{#8}%
    \fi
   \fi
  \fi
  \ifwasdraft\draft=1\else\draft=0\fi{}%
 }%
\def\TEXUX#1{"texux"}

\long\def\QQQ#1#2{%
     \long\expandafter\def\csname#1\endcsname{#2}}%
\@ifundefined{QTP}{\def\QTP#1{}}{}
\@ifundefined{Qlb}{}{}
\@ifundefined{Qlt}{}{}
\long\def\QQA#1#2{}%
\def\QTR#1#2{{\csname#1\endcsname #2}}
\def\EXPAND#1[#2]#3{}%
\def\NOEXPAND#1[#2]#3{}%
\def\LaTeXparent#1{}%
\def\ChildStyles#1{}%
\def\ChildDefaults#1{}%
\def\QTagDef#1#2#3{}%
\def\QQfnmark#1{\footnotemark}

\@ifundefined{INDEX}{\def\INDEX#1#2{}{}}{}%
\@ifundefined{SUBINDEX}{\def\SUBINDEX#1#2#3{}{}{}}{}%
\def\initial#1{\bigbreak{\raggedright\large\bf #1}\kern 2\p@
   \penalty3000}%
\@ifundefined{ZZZ}{}{\makeatletter\input gnuindex.sty\makeatother\makeindex\makeatletter}%
\@ifundefined{abstract}{%
 \def\abstract{%
  \if@twocolumn
   \section*{Abstract (Not appropriate in this style!)}%
   \else \small 
   \begin{center}{\bf Abstract\vspace{-.5em}\vspace{\z@}}\end{center}%
   \quotation 
   \fi
  }%
 }{%
 }%
\@ifundefined{endabstract}{\def\endabstract
  {\if@twocolumn\else\endquotation\fi}}{}%
\@ifundefined{maketitle}{\def\maketitle#1{}}{}%
\@ifundefined{affiliation}{\def\affiliation#1{}}{}%
\@ifundefined{proof}{}{}%
\@ifundefined{endproof}{}{}%
\@ifundefined{newfield}{\def\newfield#1#2{}}{}%
\@ifundefined{chapter}{\def\chapter#1{\par(Chapter head:)#1\par }%
 \newcount\c@chapter}{}%
\@ifundefined{part}{\def\part#1{\par(Part head:)#1\par }}{}%
\@ifundefined{section}{\def\section#1{\par(Section head:)#1\par }}{}%
\@ifundefined{subsection}{\def\subsection#1%
 {\par(Subsection head:)#1\par }}{}%
\@ifundefined{subsubsection}{\def\subsubsection#1%
 {\par(Subsubsection head:)#1\par }}{}%
\@ifundefined{paragraph}{\def\paragraph#1%
 {\par(Subsubsubsection head:)#1\par }}{}%
\@ifundefined{subparagraph}{\def\subparagraph#1%
 {\par(Subsubsubsubsection head:)#1\par }}{}%
\@ifundefined{therefore}{}{}%
\@ifundefined{backepsilon}{}{}%
\@ifundefined{yen}{}{}%
\@ifundefined{registered}{%
   \def\registered{\relax\ifmmode{}\r@gistered
                    \else$\m@th\r@gistered$\fi}%
 \def\r@gistered{^{\ooalign
  {\hfil\raise.07ex\hbox{$\scriptstyle\rm\text{R}$}\hfil\crcr
  \mathhexbox20D}}}}{}%
\@ifundefined{Eth}{}{}%
\@ifundefined{eth}{}{}%
\@ifundefined{Thorn}{}{}%
\@ifundefined{thorn}{}{}%
\@ifundefined{degree}{}{}%
\newdimen\theight
\def\Column{%
 \vadjust{\setbox\z@=\hbox{\scriptsize\quad\quad tcol}%
  \theight=\ht\z@\advance\theight by \dp\z@\advance\theight by \lineskip
  \kern -\theight \vbox to \theight{%
   \rightline{\rlap{\box\z@}}%
   \vss
   }%
  }%
 }%
\def\qed{%
 \ifhmode\unskip\nobreak\fi\ifmmode\ifinner\else\hskip5\p@\fi\fi
 \hbox{\hskip5\p@\vrule width4\p@ height6\p@ depth1.5\p@\hskip\p@}%
 }%
\def\miss{\hbox{\vrule height2\p@ width 2\p@ depth\z@}}%
\def\tcol#1{{\baselineskip=6\p@ \vcenter{#1}} \Column}  %
\def\newfmtname{LaTeX2e}
\def\chkcompat{%
   \if@compatibility
   \else
     \usepackage{latexsym}
   \fi
}

\ifx\fmtname\newfmtname
  \DeclareOldFontCommand{\rm}{\normalfont\rmfamily}{\mathrm}
  \DeclareOldFontCommand{\sf}{\normalfont\sffamily}{\mathsf}
  \DeclareOldFontCommand{\tt}{\normalfont\ttfamily}{\mathtt}
  \DeclareOldFontCommand{\bf}{\normalfont\bfseries}{\mathbf}
  \DeclareOldFontCommand{\it}{\normalfont\itshape}{\mathit}
  \DeclareOldFontCommand{\sl}{\normalfont\slshape}{\@nomath\sl}
  \DeclareOldFontCommand{\sc}{\normalfont\scshape}{\@nomath\sc}
  \chkcompat
\fi

\def\alpha{\Greekmath 010B }%
\def\beta{\Greekmath 010C }%
\def\gamma{\Greekmath 010D }%
\def\delta{\Greekmath 010E }%
\def\epsilon{\Greekmath 010F }%
\def\zeta{\Greekmath 0110 }%
\def\eta{\Greekmath 0111 }%
\def\theta{\Greekmath 0112 }%
\def\iota{\Greekmath 0113 }%
\def\kappa{\Greekmath 0114 }%
\def\lambda{\Greekmath 0115 }%
\def\mu{\Greekmath 0116 }%
\def\nu{\Greekmath 0117 }%
\def\xi{\Greekmath 0118 }%
\def\pi{\Greekmath 0119 }%
\def\rho{\Greekmath 011A }%
\def\sigma{\Greekmath 011B }%
\def\tau{\Greekmath 011C }%
\def\upsilon{\Greekmath 011D }%
\def\phi{\Greekmath 011E }%
\def\chi{\Greekmath 011F }%
\def\psi{\Greekmath 0120 }%
\def\omega{\Greekmath 0121 }%
\def\varepsilon{\Greekmath 0122 }%
\def\vartheta{\Greekmath 0123 }%
\def\varpi{\Greekmath 0124 }%
\def\varrho{\Greekmath 0125 }%
\def\varsigma{\Greekmath 0126 }%
\def\varphi{\Greekmath 0127 }%

\def\nabla{\Greekmath 0272 }

\def\Greekmath#1#2#3#4{%
    \if@compatibility
        \ifnum\mathgroup=\symbold
           \mathchoice{\mbox{\boldmath$\displaystyle\mathchar"#1#2#3#4$}}%
                      {\mbox{\boldmath$\textstyle\mathchar"#1#2#3#4$}}%
                      {\mbox{\boldmath$\scriptstyle\mathchar"#1#2#3#4$}}%
                      {\mbox{\boldmath$\scriptscriptstyle\mathchar"#1#2#3#4$}}%
        \else
           \mathchar"#1#2#3#4%
        \fi 
    \else 
        \ifnum\mathgroup=5 
           \mathchoice{\mbox{\boldmath$\displaystyle\mathchar"#1#2#3#4$}}%
                      {\mbox{\boldmath$\textstyle\mathchar"#1#2#3#4$}}%
                      {\mbox{\boldmath$\scriptstyle\mathchar"#1#2#3#4$}}%
                      {\mbox{\boldmath$\scriptscriptstyle\mathchar"#1#2#3#4$}}%
        \else
           \mathchar"#1#2#3#4%
        \fi     	    
	  \fi}

\newif\ifGreekBold  \GreekBoldfalse
\let\SAVEPBF=\pbf
\def\pbf{\GreekBoldtrue\SAVEPBF}%

\@ifundefined{theorem}{\newtheorem{theorem}{Theorem}}{}
\@ifundefined{lemma}{}{}
\@ifundefined{corollary}{}{}
\@ifundefined{conjecture}{}{}
\@ifundefined{proposition}{\newtheorem{proposition}[theorem]{Proposition}}{}
\@ifundefined{axiom}{}{}
\@ifundefined{remark}{}{}
\@ifundefined{example}{}{}
\@ifundefined{exercise}{}{}
\@ifundefined{definition}{}{}

\@ifundefined{mathletters}{%
  \newcounter{equationnumber}  
  \def\mathletters{%
     \addtocounter{equation}{1}
     \edef\@currentlabel{\theequation}%
     \setcounter{equationnumber}{\c@equation}
     \setcounter{equation}{0}%
     \edef\theequation{\@currentlabel\noexpand\alph{equation}}%
  }
  
}{}

\@ifundefined{BibTeX}{%
    \def\BibTeX{{\rm B\kern-.05em{\sc i\kern-.025em b}\kern-.08em
                 T\kern-.1667em\lower.7ex\hbox{E}\kern-.125emX}}}{}%
\@ifundefined{AmS}%
    {\def\AmS{{\protect\usefont{OMS}{cmsy}{m}{n}%
                A\kern-.1667em\lower.5ex\hbox{M}\kern-.125emS}}}{}%
\@ifundefined{AmSTeX}{}{}%
\ifx\ds@amstex\relax
\makeatother\endinput
\else
   \@ifpackageloaded{amstex}%
      {\message{amstex already loaded}\makeatother\endinput}
      {}
\fi
\let\DOTSI\relax
\def\RIfM@{\relax\ifmmode}%
\def\FN@{\futurelet\next}%
\newcount\intno@
\def\iint{\DOTSI\intno@\tw@\FN@\ints@}%
\def\iiint{\DOTSI\intno@\thr@@\FN@\ints@}%
\def\iiiint{\DOTSI\intno@4 \FN@\ints@}%
\def\idotsint{\DOTSI\intno@\z@\FN@\ints@}%
\def\ints@{\findlimits@\ints@@}%
\newif\iflimtoken@
\newif\iflimits@
\def\findlimits@{\limtoken@true\ifx\next\limits\limits@true
 \else\ifx\next\nolimits\limits@false\else
 \limtoken@false\ifx\ilimits@\nolimits\limits@false\else
 \ifinner\limits@false\else\limits@true\fi\fi\fi\fi}%
\def\multint@{\int\ifnum\intno@=\z@\intdots@                          
 \else\intkern@\fi                                                    
 \ifnum\intno@>\tw@\int\intkern@\fi                                   
 \ifnum\intno@>\thr@@\int\intkern@\fi                                 
 \int}
\def\multintlimits@{\intop\ifnum\intno@=\z@\intdots@\else\intkern@\fi
 \ifnum\intno@>\tw@\intop\intkern@\fi
 \ifnum\intno@>\thr@@\intop\intkern@\fi\intop}%
\def\intic@{%
    \mathchoice{\hskip.5em}{\hskip.4em}{\hskip.4em}{\hskip.4em}}%
\def\negintic@{\mathchoice
 {\hskip-.5em}{\hskip-.4em}{\hskip-.4em}{\hskip-.4em}}%
\def\ints@@{\iflimtoken@                                              
 \def\ints@@@{\iflimits@\negintic@
   \mathop{\intic@\multintlimits@}\limits                             
  \else\multint@\nolimits\fi                                          
  \eat@}
 \else                                                                
 \def\ints@@@{\iflimits@\negintic@
  \mathop{\intic@\multintlimits@}\limits\else
  \multint@\nolimits\fi}\fi\ints@@@}%
\def\intkern@{\mathchoice{\!\!\!}{\!\!}{\!\!}{\!\!}}%
\def\plaincdots@{\mathinner{\cdotp\cdotp\cdotp}}%
\def\intdots@{\mathchoice{\plaincdots@}%
 {{\cdotp}\mkern1.5mu{\cdotp}\mkern1.5mu{\cdotp}}%
 {{\cdotp}\mkern1mu{\cdotp}\mkern1mu{\cdotp}}%
 {{\cdotp}\mkern1mu{\cdotp}\mkern1mu{\cdotp}}}%
\def\RIfM@{\relax\protect\ifmmode}
\def\text{\RIfM@\expandafter\text@\else\expandafter\mbox\fi}
\let\nfss@text\text
\def\text@#1{\mathchoice
   {\textdef@\displaystyle\f@size{#1}}%
   {\textdef@\textstyle\tf@size{\firstchoice@false #1}}%
   {\textdef@\textstyle\sf@size{\firstchoice@false #1}}%
   {\textdef@\textstyle \ssf@size{\firstchoice@false #1}}%
   \glb@settings}

\def\textdef@#1#2#3{\hbox{{%
                    \everymath{#1}%
                    \let\f@size#2\selectfont
                    #3}}}
\newif\iffirstchoice@
\firstchoice@true
\def\Let@{\relax\iffalse{\fi\let\\=\cr\iffalse}\fi}%
\def\vspace@{\def\vspace##1{\crcr\noalign{\vskip##1\relax}}}%
\def\multilimits@{\bgroup\vspace@\Let@
 \baselineskip\fontdimen10 \scriptfont\tw@
 \advance\baselineskip\fontdimen12 \scriptfont\tw@
 \lineskip\thr@@\fontdimen8 \scriptfont\thr@@
 \lineskiplimit\lineskip
 \vbox\bgroup\ialign\bgroup\hfil$\m@th\scriptstyle{##}$\hfil\crcr}%
\def\Sb{_\multilimits@}%
\def\endSb{\crcr\egroup\egroup\egroup}%
\def\Sp{^\multilimits@}%

\newdimen\ex@
\def\rightarrowfill@#1{$#1\m@th\mathord-\mkern-6mu\cleaders
 \hbox{$#1\mkern-2mu\mathord-\mkern-2mu$}\hfill
 \mkern-6mu\mathord\rightarrow$}%
\def\leftarrowfill@#1{$#1\m@th\mathord\leftarrow\mkern-6mu\cleaders
 \hbox{$#1\mkern-2mu\mathord-\mkern-2mu$}\hfill\mkern-6mu\mathord-$}%
\def\leftrightarrowfill@#1{$#1\m@th\mathord\leftarrow
\mkern-6mu\cleaders
 \hbox{$#1\mkern-2mu\mathord-\mkern-2mu$}\hfill
 \mkern-6mu\mathord\rightarrow$}%
\def\overrightarrow{\mathpalette\overrightarrow@}%
\def\overrightarrow@#1#2{\vbox{\ialign{##\crcr\rightarrowfill@#1\crcr
 \noalign{\kern-\ex@\nointerlineskip}$\m@th\hfil#1#2\hfil$\crcr}}}%

\def\overleftarrow{\mathpalette\overleftarrow@}%
\def\overleftarrow@#1#2{\vbox{\ialign{##\crcr\leftarrowfill@#1\crcr
 \noalign{\kern-\ex@\nointerlineskip}$\m@th\hfil#1#2\hfil$\crcr}}}%
\def\overleftrightarrow{\mathpalette\overleftrightarrow@}%
\def\overleftrightarrow@#1#2{\vbox{\ialign{##\crcr
   \leftrightarrowfill@#1\crcr
 \noalign{\kern-\ex@\nointerlineskip}$\m@th\hfil#1#2\hfil$\crcr}}}%
\def\underrightarrow{\mathpalette\underrightarrow@}%
\def\underrightarrow@#1#2{\vtop{\ialign{##\crcr$\m@th\hfil#1#2\hfil
  $\crcr\noalign{\nointerlineskip}\rightarrowfill@#1\crcr}}}%

\def\underleftarrow{\mathpalette\underleftarrow@}%
\def\underleftarrow@#1#2{\vtop{\ialign{##\crcr$\m@th\hfil#1#2\hfil
  $\crcr\noalign{\nointerlineskip}\leftarrowfill@#1\crcr}}}%
\def\underleftrightarrow{\mathpalette\underleftrightarrow@}%
\def\underleftrightarrow@#1#2{\vtop{\ialign{##\crcr$\m@th
  \hfil#1#2\hfil$\crcr
 \noalign{\nointerlineskip}\leftrightarrowfill@#1\crcr}}}%

\def\qopnamewl@#1{\mathop{\operator@font#1}\nlimits@}
\let\nlimits@\displaylimits
\def\setboxz@h{\setbox\z@\hbox}

\def\varlim@#1#2{\mathop{\vtop{\ialign{##\crcr
 \hfil$#1\m@th\operator@font lim$\hfil\crcr
 \noalign{\nointerlineskip}#2#1\crcr
 \noalign{\nointerlineskip\kern-\ex@}\crcr}}}}

 \def\rightarrowfill@#1{\m@th\setboxz@h{$#1-$}\ht\z@\z@
  $#1\copy\z@\mkern-6mu\cleaders
  \hbox{$#1\mkern-2mu\box\z@\mkern-2mu$}\hfill
  \mkern-6mu\mathord\rightarrow$}
\def\leftarrowfill@#1{\m@th\setboxz@h{$#1-$}\ht\z@\z@
  $#1\mathord\leftarrow\mkern-6mu\cleaders
  \hbox{$#1\mkern-2mu\copy\z@\mkern-2mu$}\hfill
  \mkern-6mu\box\z@$}

\def\projlim{\qopnamewl@{proj\,lim}}
\def\injlim{\qopnamewl@{inj\,lim}}
\def\varinjlim{\mathpalette\varlim@\rightarrowfill@}
\def\varprojlim{\mathpalette\varlim@\leftarrowfill@}
\def\varliminf{\mathpalette\varliminf@{}}
\def\varliminf@#1{\mathop{\underline{\vrule\@depth.2\ex@\@width\z@
   \hbox{$#1\m@th\operator@font lim$}}}}
\def\varlimsup{\mathpalette\varlimsup@{}}
\def\varlimsup@#1{\mathop{\overline
  {\hbox{$#1\m@th\operator@font lim$}}}}

\def\QTATOP#1#2{{\textstyle {#1 \atop #2}}}%
\begingroup \catcode `|=0 \catcode `[= 1
\catcode`]=2 \catcode `\{=12 \catcode `\}=12
\catcode`\\=12 
|gdef|@alignverbatim#1\end{align}[#1|end[align]]
|gdef|@salignverbatim#1\end{align*}[#1|end[align*]]

|gdef|@alignatverbatim#1\end{alignat}[#1|end[alignat]]
|gdef|@salignatverbatim#1\end{alignat*}[#1|end[alignat*]]

|gdef|@xalignatverbatim#1\end{xalignat}[#1|end[xalignat]]
|gdef|@sxalignatverbatim#1\end{xalignat*}[#1|end[xalignat*]]

|gdef|@gatherverbatim#1\end{gather}[#1|end[gather]]
|gdef|@sgatherverbatim#1\end{gather*}[#1|end[gather*]]

|gdef|@gatherverbatim#1\end{gather}[#1|end[gather]]
|gdef|@sgatherverbatim#1\end{gather*}[#1|end[gather*]]

|gdef|@multilineverbatim#1\end{multiline}[#1|end[multiline]]
|gdef|@smultilineverbatim#1\end{multiline*}[#1|end[multiline*]]

|gdef|@arraxverbatim#1\end{arrax}[#1|end[arrax]]
|gdef|@sarraxverbatim#1\end{arrax*}[#1|end[arrax*]]

|gdef|@tabulaxverbatim#1\end{tabulax}[#1|end[tabulax]]
|gdef|@stabulaxverbatim#1\end{tabulax*}[#1|end[tabulax*]]

|endgroup

\def\align{\@verbatim \frenchspacing\@vobeyspaces \@alignverbatim
You are using the "align" environment in a style in which it is not defined.}

\@namedef{align*}{\@verbatim\@salignverbatim
You are using the "align*" environment in a style in which it is not defined.}
\expandafter\let\csname endalign*\endcsname =\endtrivlist

\def\alignat{\@verbatim \frenchspacing\@vobeyspaces \@alignatverbatim
You are using the "alignat" environment in a style in which it is not defined.}

\@namedef{alignat*}{\@verbatim\@salignatverbatim
You are using the "alignat*" environment in a style in which it is not defined.}
\expandafter\let\csname endalignat*\endcsname =\endtrivlist

\def\xalignat{\@verbatim \frenchspacing\@vobeyspaces \@xalignatverbatim
You are using the "xalignat" environment in a style in which it is not defined.}

\@namedef{xalignat*}{\@verbatim\@sxalignatverbatim
You are using the "xalignat*" environment in a style in which it is not defined.}
\expandafter\let\csname endxalignat*\endcsname =\endtrivlist

\def\gather{\@verbatim \frenchspacing\@vobeyspaces \@gatherverbatim
You are using the "gather" environment in a style in which it is not defined.}

\@namedef{gather*}{\@verbatim\@sgatherverbatim
You are using the "gather*" environment in a style in which it is not defined.}
\expandafter\let\csname endgather*\endcsname =\endtrivlist

\def\multiline{\@verbatim \frenchspacing\@vobeyspaces \@multilineverbatim
You are using the "multiline" environment in a style in which it is not defined.}

\@namedef{multiline*}{\@verbatim\@smultilineverbatim
You are using the "multiline*" environment in a style in which it is not defined.}
\expandafter\let\csname endmultiline*\endcsname =\endtrivlist

\def\arrax{\@verbatim \frenchspacing\@vobeyspaces \@arraxverbatim
You are using a type of "array" construct that is only allowed in AmS-LaTeX.}

\def\tabulax{\@verbatim \frenchspacing\@vobeyspaces \@tabulaxverbatim
You are using a type of "tabular" construct that is only allowed in AmS-LaTeX.}

\@namedef{arrax*}{\@verbatim\@sarraxverbatim
You are using a type of "array*" construct that is only allowed in AmS-LaTeX.}
\expandafter\let\csname endarrax*\endcsname =\endtrivlist

\@namedef{tabulax*}{\@verbatim\@stabulaxverbatim
You are using a type of "tabular*" construct that is only allowed in AmS-LaTeX.}
\expandafter\let\csname endtabulax*\endcsname =\endtrivlist

\def\@@eqncr{\let\@tempa\relax
    \ifcase\@eqcnt \def\@tempa{& & &}\or \def\@tempa{& &}%
      \else \def\@tempa{&}\fi
     \@tempa
     \if@eqnsw
        \iftag@
           \@taggnum
        \else
           \@eqnnum\stepcounter{equation}%
        \fi
     \fi
     \global\tag@false
     \global\@eqnswtrue
     \global\@eqcnt\z@\cr}

 \def\endequation{%
     \ifmmode\ifinner 
      \iftag@
        \addtocounter{equation}{-1} 
        $\hfil
           \displaywidth\linewidth\@taggnum\egroup \endtrivlist
        \global\tag@false
        \global\@ignoretrue   
      \else
        $\hfil
           \displaywidth\linewidth\@eqnnum\egroup \endtrivlist
        \global\tag@false
        \global\@ignoretrue 
      \fi
     \else   
      \iftag@
        \addtocounter{equation}{-1} 
        \eqno \hbox{\@taggnum}
        \global\tag@false%
        $$\global\@ignoretrue
      \else
        \eqno \hbox{\@eqnnum}
        $$\global\@ignoretrue
      \fi
     \fi\fi
 } 

 \newif\iftag@ \tag@false
 
 \def\tag{\@ifnextchar*{\@tagstar}{\@tag}}
 \def\@tag#1{%
     \global\tag@true
     \global\def\@taggnum{(#1)}}
 \def\@tagstar*#1{%
     \global\tag@true
     \global\def\@taggnum{#1}%
}

\makeatother

\begin{document}

\begin{flushright}
\setlength{\baselineskip}{3ex}
\#HUTP-97/A063\\10/97\end{flushright}

\begin{center}
{\Large \textbf{{Singularities of massless planar diagrams, large-$N_c$
mesons in $3+1$ dimensions, and the 't Hooft model} }{\normalsize \textrm{\\ %
\vspace{12pt} Dean Lee\footnote{Supported by
the National Science Foundation under Grant \#PHY-9218167 and the Fannie and
John Hertz Foundation.}\\ Harvard University\\ Cambridge, MA 02138\\ %
\vspace{24pt} 
\small
\parbox{360pt}{We study the singular Landau 
surfaces of planar diagrams contributing to scattering of a massless quark and antiquark 
in $3+1$ dimensions. 
In particular, we look at singularities which remain after integration with respect to the 
various angular degrees of freedom. We derive a general relation between these singularities and
the singularities of quark-antiquark scattering in $1+1$ dimensions. We then classify all 
Landau surfaces 
of the $1+1$ dimensional system. Combining these results, we deduce that the singular 
surfaces of the angle-integrated $3+1$ dimensional amplitude must satisfy at least one of three 
conditions, which we call the planar light-cone conditions. We discuss the extension of 
our results to non-perturbative processes by means of the non-perturbative 
operator product expansion.  Our findings offer new 
insights into the connection between the 't Hooft 
model and large-$N_c$ mesons in $3+1$ dimensions and may prove useful in studies of 
confinement in relativistic meson systems.}
\normalsize\\ \vspace{10pt} }}}
\end{center}

\setcounter{footnote}{0}

\section{Overview}

The 't Hooft model \cite{thooft}, which describes large-$N_c$ mesons in $1+1$
dimensions, is a popular starting point for studies of quark confinement in
mesons. The combination of large-$N_c$ and the low number of dimensions make
it possible to calculate the Bethe-Salpeter equation exactly. Also the
Bethe-Salpeter equation turns out to be relatively simple, and it is
straightforward to compute the bound state wavefunctions numerically. For
these reasons the 't Hooft model serves as a useful concrete example of
confinement and relativistic bound states. Unfortunately, very little is
known about the relation between confinement in the 't Hooft model and
confinement in the real world. In this paper we clarify some aspects of this
relation.

In our analysis we compare large-$N_c$ massless QCD in $1+1$ dimensions and $%
3+1$ dimensions in terms of their corresponding Bethe-Salpeter integral
kernels. Since confinement is associated with the infrared behavior of the
interaction, our analysis will center on the momentum-space singularities of
the Bethe-Salpeter kernel. The $3+1$ dimensional system carries angular
degrees of freedom not in the $1+1$ dimensional system, and so we consider,
in the $3+1$ dimensional case, the form of the Bethe-Salpeter kernel
restricted to incoming and outgoing states with fixed intrinsic spin and
orbital angular momentum. In esssence we integrate the $3+1$ dimensional
kernel, along with terms from the spherical harmonics, with respect to the
angular degrees freedom.

Our work leads to two main results. The first result is that the
singularities of the angle-integrated $3+1$ dimensional kernel, calculated
to any finite number of loops, satisfy the same Landau equations that
describe the singularities of the $1+1$ dimensional kernel. In plain terms
this means that the singularities of the two kernels occur at the same place
in momentum-space. The only difference is the behavior near the singularity
(e.g., the two kernels may diverge with different coefficients and critical
exponents). As we will see, planarity of large-$N_c$ diagrams plays an
important role in establishing this relation. The second result is the
classification of all the singular surfaces of the $1+1$ dimensional kernel
at any finite number of loops. By the first result this is also a
classification of the singular surfaces of the angle-integrated $3+1$
dimensional kernel.

The organization of the paper is as follows. In the section 2, we provide
background material on the Chisholm method for computing Feynman amplitudes.
In section 3 we analyze the singularities corresponding with massless planar
diagrams in $3+1$ dimensions and derive the first result. In section 4
examine the singularities of massless planar diagrams in $1+1$ dimensions
and classify all possible singular surfaces, our second result. In section 5
we discuss how to apply our methods to the non-perturbative operator product
expansion. In section 6 we summarize our findings and discuss their
significance.

\section{Chisholm method}

Let $\frak{g}$ be a 1PI diagram for scattering of a massless quark and
antiquark, with momentum assignments as shown in Figure 1. We define $%
g_{(q,q^{\prime });\varepsilon }^{ab;cd}(p_0)$ to be the $\varepsilon $%
-regulated amplitude of $\frak{g}$ in the rest frame of $p$, where $%
\varepsilon $ is the infinitesimal positive parameter used to regulate the
singularities of the particle propagators. Since gauge-fixing affects the
numerator of the gauge propagator rather than the singularities associated
with the denominator, our method of gauge fixing will not affect our
analysis. To avoid unnecessary abstract language, however, we will work in a
covariant gauge. Ultraviolet regularization and renormalization also do not
play a role in the analysis. We know this because the theory we are
considering is renormalizable, and all counterterms are proportional to the
original vertex interactions. Diagrams with counterterms and diagrams
without counterterms therefore generate the same momentum-space
singularities.

Let $n$ be the number of internal lines and $L$ be the number of loops in $%
\frak{g}$. Each gauge propagator has terms proportional to $\frac{g^{\mu \nu
}}{k_j^2+i\varepsilon }$ and $\frac{k_j^\mu k_j^\nu }{\left(
k_j^2+i\varepsilon \right) ^2}$. We treat these terms separately when
combining denominators by the Feynman parameter method. If $n^{\prime }$ is
the number of gauge propagators there are $2^{n^{\prime }}$ choices for the
various terms, and we use the index $f=1,\cdots $ $,2^{n^{\prime }}$ to
denote a particular choice. For each $f$ let $m_f$ be the total number of
terms proportional to $\frac{k_j^\mu k_j^\nu }{\left( k_j^2+i\varepsilon
\right) ^2}$. We can write the amplitude as 
\begin{equation}
g_{(q,q^{\prime });\varepsilon
}^{ab;cd}(p_0)=\sum\limits_{f=1}^{2^{n^{\prime }}}\int\limits_{\alpha \in
\sigma ^n}\frac{d^n\alpha \,d^{4L}l\,P_f^{ab;cd}(p_0,q,q^{\prime },l,\alpha )%
}{\left[ \sum\limits_{j=1}^n\alpha _jk_j^2+i\varepsilon \right] ^{n+m_f}},
\label{one}
\end{equation}
where each $P_f^{ab;cd}(p_0,q,q^{\prime },l,\alpha )$ is a polynomial, $%
\sigma ^n=\left\{ \alpha \left| \alpha _j\geq 0,\sum\limits_j\alpha
_j=1\right. \right\} $, and for each internal line $j=1,\cdots ,n,$ $k_j$ is
the momentum flowing through line $j.$

Our purpose is to study the singular points of $g_{(q,q^{\prime
});\varepsilon }^{ab;cd}(p_0)$. In particular, we are interested in the
singular surfaces which remain after integration with respect to the angular
variables $\hat{q}$ and $\hat{q}^{\prime }$. Although the polynomial $%
P_f^{ab;cd}$ affects the behavior of $g_{(q,q^{\prime });\varepsilon
}^{ab;cd}(p_0)$ near the singular surfaces, the structure of the surfaces is
determined by the zero-set of the denominator, 
\begin{equation}
\sum\limits_{j=1}^n\alpha _jk_j^2+i\varepsilon .  \label{btrrt}
\end{equation}
In some sense the singular points of $g_{(q,q^{\prime });\varepsilon
}^{ab;cd}(p_0)$ are a subset of the singular points of the function 
\begin{equation}
g_\varepsilon (p_0,q,q^{\prime })=\int\limits_{\alpha \in \sigma ^n}\frac{%
d^n\alpha \,d^{4L}l\,}{\left[ \sum\limits_{j=1}^n\alpha _jk_j^2+i\varepsilon
\right] ^n}\propto \int d^{4L}l\prod\limits_{j=1}^n\frac{\,1}{%
k_j^2+i\varepsilon }.  \label{rewy}
\end{equation}
In more precise terms, each singular point of $g_{(q,q^{\prime
});\varepsilon }^{ab;cd}(p_0)$ is a solution of the Landau equations%
\footnote{%
The Landau equations are written out in equations (\ref{land}) and (\ref{dau}%
) of the next section.} of $g_\varepsilon (p_0,q,q^{\prime })$. These two
statements can differ when a solution of the Landau equations of $%
g_\varepsilon (p_0,q,q^{\prime })$ is a regular point of $g_\varepsilon
(p_0,q,q^{\prime })$ as a result of accidental cancellation.

Throughout the rest of this section, we consider the singular points of 
\begin{equation}
g_{\vec{\varepsilon}}(p_0,q,q^{\prime })=\int\limits_{\alpha \in \sigma ^n}%
\frac{d^n\alpha \,d^{4L}l\,}{\left[ \sum\limits_{j=1}^n\alpha
_j(k_j^2+i\varepsilon _j)\right] ^n}\propto \int d^{4L}l\prod\limits_{j=1}^n%
\frac{\,1}{k_j^2+i\varepsilon _j}.  \label{mrtwe}
\end{equation}
For convenience later on we have generalized our infinitesimal parameter $%
\varepsilon $ to a vector of infinitesimal parameters $\vec{\varepsilon}%
=\left\{ \varepsilon _1,\cdots ,\varepsilon _n\right\} .$ For each loop $i$
and internal line $j$, let 
\begin{equation}
Z_{ij}=\left\{ 
\begin{array}{l}
1\text{ if loop }i\text{ contains line }j\text{ with orientation }k_j \\ 
-1\text{ if loop }i\text{ contains line }j\text{ with orientation }-k_j \\ 
0\text{ if loop }i\text{ does not contain line }j\text{.}
\end{array}
\right.  \label{two}
\end{equation}
We can perform the loop integrations \cite{chis} to get 
\begin{equation}
g_{\vec{\varepsilon}}(p_0,q,q^{\prime })=\int\limits_{\alpha \in \sigma ^n}%
\frac{d^n\alpha \,\left[ C(\alpha )\right] ^{n-2L-2}}{\left[ \sum\limits\Sb %
h=p^2,q^2,q^{\prime 2},  \\ p\cdot q,p\cdot q^{\prime },q\cdot q^{\prime } 
\endSb hD_h(\alpha )+iC(\alpha )\sum\limits_{j=1}^n\alpha _j\varepsilon
_j\right] ^{n-2L}},  \label{three}
\end{equation}
where each $D_h(\alpha )$ is a homogeneous polynomial of degree $L+1$, and $%
C(\alpha )$ is a homogeneous polynomial of degree $L$. We have dropped
various constant factors which will be irrelevant to our discussion. $%
C(\alpha )$ can be written in determinant form:$\footnote{%
See \cite{tiktop}.}$%
\begin{equation}
C(\alpha )=\det \mathbf{C}=\left| 
\begin{array}{ccc}
a_{1,1} & \cdots & a_{1,L} \\ 
\vdots & \ddots & \vdots \\ 
a_{L,1} & \cdots & a_{L,L}
\end{array}
\right|  \label{four}
\end{equation}
where $a_{i,i^{\prime }}=$ $\sum\limits_j\alpha _jZ_{ij}Z_{i^{\prime }j}$.
For each internal line $j=1,\cdots ,n,$ let $\dot{k}_j$ be the momentum
flowing through line $j$ when all independent loop momenta are set to zero.
We can write $\sum\limits\Sb h=p^2,q^2,q^{\prime 2},  \\ p\cdot q,p\cdot
q^{\prime },q\cdot q^{\prime }  \endSb hD_h(\alpha )$ as a symbolic
determinant: 
\begin{equation}
\sum\limits\Sb h=p^2,q^2,q^{\prime 2},  \\ p\cdot q,p\cdot q^{\prime
},q\cdot q^{\prime }  \endSb hD_h(\alpha )=\left| 
\begin{array}{cccc}
a_{1,1} & \cdots & a_{1,L} & a_{1,L+1} \\ 
\vdots & \ddots & \vdots & \vdots \\ 
a_{L,1} & \cdots & a_{L,L} & a_{L,L+1} \\ 
a_{L+1,1} & \cdots & a_{L+1,L} & a_{L+1,L+1}
\end{array}
\right|  \label{five}
\end{equation}
where for $1\leq i\leq L$, $a_{i,L+1}=a_{L+1,i}=$ $\sum\limits_jZ_{ij}\alpha
_j\dot{k}_j^\mu $, and $a_{L+1,L+1}=\sum\limits_{i,j}(Z_{ij})^2\alpha _j\dot{%
k}_j^2$. In our definition of the symbolic determinant, we identify the
product $a_{i,L+1}a_{i^{\prime },L+1},$ for $1\leq i,i^{\prime }\leq L$,
with the corresponding Lorentz contraction: 
\begin{equation}
\left( \sum\limits_jZ_{ij}\alpha _j\dot{k}_j^\mu \right) \left(
\sum\limits_jZ_{i^{^{\prime }}j}\alpha _j\dot{k}_{j\mu }\right) .
\label{six}
\end{equation}
Let us consider the angular integral 
\begin{equation}
G_{\vec{\varepsilon}}\left( \QTATOP{p_0,q_0,q_0^{\prime }}{\left| \vec{q}%
\right| ,\left| \vec{q}^{\text{ }\prime }\right| }\right) =\frac 1{8\pi
^2}\int d\cos \theta _{\hat{q},\hat{z}}\,d\cos \theta _{\hat{q},\hat{q}%
^{\prime }}\,d\phi _{\hat{q},\hat{z}}\,d\phi _{\hat{q},\hat{q}^{\prime }}\,%
\vec{q}^2\vec{q}^{\text{ }\prime 2}g_{\vec{\varepsilon}}(p_0,q,q^{\prime }),
\label{seven}
\end{equation}
where $\theta _{\hat{q},\hat{z}}$ is the angle between $\hat{q}$ and the $z$%
-axis, $\phi _{\hat{q},\hat{z}}$ is the azimuthal angle of $\hat{q}$ about
the $z$-axis, $\theta _{\hat{q},\hat{q}^{\prime }}$ is the angle between $%
\hat{q}$ and $\hat{q}^{\prime }$, $\phi _{\hat{q},\hat{q}^{\prime }}$ is the
azimuthal angle of $\hat{q}$ about $\hat{q}^{\prime }$. Since we are in the
rest frame where $\vec{p}=0$, all Lorentz invariants in the denominator of (%
\ref{three}) are independent of these angular variables with the exception
of $q\cdot q^{\prime }$, which is given by 
\begin{equation}
q\cdot q^{\prime }=q_0q_0^{\prime }-\left| \vec{q}\right| \left| \vec{q}^{%
\text{ }\prime }\right| \cos \theta _{\hat{q},\hat{q}^{\prime }}\text{.}
\label{wrwt}
\end{equation}
We can rewrite (\ref{seven}) as 
\begin{equation}
G_{\vec{\varepsilon}}=\,\int\limits_{\alpha \in \sigma ^n}d^n\alpha \int 
\frac{d\cos \theta _{\hat{q},\hat{q}^{\prime }}\left[ C(\alpha )\right]
^{n-2L-2}\vec{q}^2\vec{q}^{\text{ }\prime 2}}{\left[ \sum\limits\Sb %
h=p^2,q^2,q^{\prime 2},  \\ p\cdot q,p\cdot q^{\prime },q\cdot q^{\prime } 
\endSb hD_h(\alpha )+iC(\alpha )\sum\limits_{j=1}^n\alpha _j\varepsilon
_j\right] ^{n-2L}}.  \label{eew}
\end{equation}
Integrating with respect to $\cos \theta _{\hat{q},\hat{q}^{\prime }}$, we
find $G_{\vec{\varepsilon}}$ equals 
\begin{equation}
\sum\limits_{\lambda =-1,+1}\int\limits_{\alpha \in \sigma ^n}\frac{%
d^n\alpha \,\frac 1{(n-2L-1)D_{q\cdot q^{\prime }}(\alpha )}\lambda \left[
C(\alpha )\right] ^{n-2L-2}\left| \vec{q}\right| \left| \vec{q}^{\text{ }%
\prime }\right| }{\left[ \sum\limits\Sb h=p^2,q^2,q^{\prime 2},  \\ p\cdot
q,p\cdot q^{\prime }  \endSb hD_h(\alpha )+\left( q_0q_0^{\prime }-\lambda
\left| \vec{q}\right| \left| \vec{q}^{\text{ }\prime }\right| \right)
D_{q\cdot q^{\prime }}(\alpha )+iC(\alpha )\sum\limits_{j=1}^n\alpha
_j\varepsilon _j\right] ^{n-2L-1}}  \label{eedx}
\end{equation}

\section{Planar diagrams}

Let us now restrict to the case when $\frak{g}$ is a planar diagram. As an
example, let us consider the diagram shown in Figure 2. It is convenient to
route our loop momentum variables such that each independent loop
circumscribes a region (i.e., tile) of our planar diagram. In our example we
take $k_2$ and $k_6$ to be the independent loop momenta. With this
convention $\dot{k}_j$ is then defined by setting $k_2=k_6=0$. For this
diagram, $\sum\limits\Sb h=p^2,q^2,q^{\prime 2},  \\ p\cdot q,p\cdot
q^{\prime },q\cdot q^{\prime }  \endSb hD_h(\alpha )$ is given by 
\begin{equation}
\left| 
\begin{tabular}{ccc}
$\alpha _1+\alpha _2+\alpha _3+\alpha _4$ & $-\alpha _4$ & $\alpha _3\dot{k}%
_3+\alpha _4\dot{k}_4+\alpha _1\dot{k}_1$ \\ 
$-\alpha _4$ & $\alpha _4+\alpha _5+\alpha _6+\alpha _7$ & $\alpha _7\dot{k}%
_7-\alpha _4\dot{k}_4+\alpha _5\dot{k}_5$ \\ 
$\alpha _3\dot{k}_3+\alpha _4\dot{k}_4+\alpha _1\dot{k}_1$ & $\alpha _7\dot{k%
}_7-\alpha _4\dot{k}_4+\alpha _5\dot{k}_5$ & $a_{33}$%
\end{tabular}
\right|  \label{hgew}
\end{equation}
where 
\begin{eqnarray}
2\dot{k}_1 &=&p+q^{\prime }  \label{nay} \\
2\dot{k}_3 &=&p+q  \nonumber \\
2\dot{k}_4 &=&2p  \nonumber \\
2\dot{k}_5 &=&-p+q  \nonumber \\
2\dot{k}_7 &=&-p+q^{\prime }  \nonumber
\end{eqnarray}
and 
\begin{equation}
a_{33}=\alpha _1\dot{k}_1^2+\alpha _3\dot{k}_3^2+\alpha _4\dot{k}_4^2+\alpha
_5\dot{k}_5^2+\alpha _7\dot{k}_7^2.  \label{xvbn}
\end{equation}
We find that $D_{q\cdot q^{\prime }}(\alpha )$ equals 
\begin{equation}
-\frac 12\left[ \alpha _1\alpha _3(\alpha _4+\alpha _5+\alpha _6+\alpha
_7)+\alpha _5\alpha _7(\alpha _1+\alpha _2+\alpha _3+\alpha _4)+(\alpha
_1\alpha _5+\alpha _3\alpha _7)\alpha _4\right] \text{.}  \label{ytte}
\end{equation}
There is a correspondence between the terms in $D_{q\cdot q^{\prime
}}(\alpha )$ and paths from the left-side of $\frak{g}$ to the right-side of 
$\frak{g}$.\footnote{%
This correspondence is proven in \cite{tiktop}.} We define a path to be a
continuous walk through adjacent interior regions such that no region is
entered twice. We have shown all possible left-to-right paths for $\frak{g}$
in Figures 2(a-d)$.$ For each path $\rho $, let $\left| \rho \right| $ be
the total number of internal lines crossed. We use an alternating string of $%
\alpha $'s and $l$'s$,$ 
\begin{equation}
\left\{ \alpha _{j_1(\rho )},l_{i_1(\rho )},\cdots \alpha _{j_{\left| \rho
\right| -1}(\rho )},l_{i_{\left| \rho \right| -1}(\rho )}\alpha _{j_{\left|
\rho \right| }(\rho )}\right\} ,  \label{cvbnr}
\end{equation}
to denote the sequence of internal lines crossed and loop regions entered by 
$\rho $. With each string we associate the polynomial 
\begin{equation}
\alpha _{j_1(\rho )}\alpha _{j_2(\rho )}\cdots \alpha _{j_{\left| \rho
\right| }(\rho )}C(\alpha ,i_1(\rho ),\cdots ,i_{\left| \rho \right|
-1}(\rho )),  \label{gfew}
\end{equation}
where $C(\alpha ,i_1(\rho ),\cdots ,i_{\left| \rho \right| }(\rho ))$ is the
determinant of the submatrix generated by striking out the $i_1(\rho
),\cdots ,i_{\left| \rho \right| }(\rho )$ rows and columns of the matrix $%
\mathbf{C}$ shown in equation (\ref{four})$.$ When $\left| \rho \right| -1=L$%
, we take $C(\alpha ,i_1(\rho ),\cdots ,i_{\left| \rho \right| -1}(\rho ))=1$%
. The expression for $D_{q\cdot q^{\prime }}(\alpha )$ is then 
\begin{equation}
D_{q\cdot q^{\prime }}(\alpha )=-\frac 12\sum_{\text{paths }\rho }\alpha
_{j_1(\rho )}\alpha _{j_2(\rho )}\cdots \alpha _{j_{\left| \rho \right|
}(\rho )}C(\alpha ,i_1(\rho ),\cdots ,i_{\left| \rho \right| -1}(\rho ))%
\text{.}  \label{rtds}
\end{equation}
When $\alpha _j\geq 0$ for each $j$, we find $D_{q\cdot q^{\prime }}(\alpha
)\leq 0$. In fact $D_{q\cdot q^{\prime }}(\alpha )\leq 0$ for any planar
diagram since $C(\alpha ,i_1(\rho ),\cdots ,i_{\left| \rho \right| -1}(\rho
))$ is the determinant of a positive quadratic form. Under the constraint $%
\alpha _j\geq 0$ for each $j$, we can use our correspondence with
left-to-right paths to conclude that $D_{q\cdot q^{\prime }}(\alpha )=0$ if
and only if there exists $j_1^{\prime },\cdots ,j_r^{\prime }$ such that 
\begin{equation}
\alpha _{j_1^{\prime }}=\alpha _{j_2^{\prime }}=\cdots =\alpha _{j_r^{\prime
}}=0  \label{wefw}
\end{equation}
and for every left-to-right path $\rho ,$ 
\begin{equation}
\left\{ j_1(\rho ),\cdots ,j_{\left| \rho \right| }(\rho )\right\} \cap
\left\{ j_1^{\prime },\cdots ,j_r^{\prime }\right\} \neq \emptyset \text{.}
\label{rewu}
\end{equation}

Let $-\frac 12\alpha _{j_1}\alpha _{j_2}\cdots \alpha _{j_{L+1}}$ be any
monomial appearing in $D_{q\cdot q^{\prime }}(\alpha )$. Taking successive
derivatives of $G_{\vec{\varepsilon}}$ with respect to $\varepsilon
_{j_1},\varepsilon _{j_2}$,$\cdots ,\varepsilon _{j_{L+1}}$ we find $\frac
d{d\varepsilon _{j_1}}\cdots \frac d{d\varepsilon _{j_{L+1}}}G_{\vec{%
\varepsilon}}$ is proportional to 
\begin{equation}
\sum\limits_{\lambda =-1,+1}\int\limits_{\alpha \in \sigma ^n}\frac{%
d^n\alpha \,\frac{\,\alpha _{j_1}\alpha _{j_2}\cdots \alpha _{j_{L+1}}i^{L+1}%
}{(n-2L-1)D_{q\cdot q^{\prime }}(\alpha )}\lambda \left[ C(\alpha )\right]
^{n-L-1}\left| \vec{q}\right| \left| \vec{q}^{\text{ }\prime }\right| }{%
\left[ \sum\limits\Sb h=p^2,q^2,q^{\prime 2},  \\ p\cdot q,p\cdot q^{\prime
}  \endSb hD_h(\alpha )+\left( q_0q_0^{\prime }-\lambda \left| \vec{q}%
\right| \left| \vec{q}^{\text{ }\prime }\right| \right) D_{q\cdot q^{\prime
}}(\alpha )+i\sum\limits_{j=1}^n\alpha _j\varepsilon _j\cdot C(\alpha
)\right] ^{n-L}}\text{.}  \label{geuu}
\end{equation}
We recall that $D_{q\cdot q^{\prime }}(\alpha )$ is a sum of monomials each
with negative coefficient$.$ Since one of these monomials is $-\frac
12\alpha _{j_1}\alpha _{j_2}\cdots \alpha _{j_{L+1}}$, 
\begin{equation}
D_{q\cdot q^{\prime }}(\alpha )\leq -\frac 12\alpha _{j_1}\alpha
_{j_2}\cdots \alpha _{j_{L+1}}\leq 0  \label{hgtr}
\end{equation}
for all $\alpha \in \sigma ^n$. We conclude that $\frac{\alpha _{j_1}\alpha
_{j_2}\cdots \alpha _{j_{L+1}}}{D_{q\cdot q^{\prime }}(\alpha )}$ is
analytic on the interior of $\sigma ^n$ and bounded on $\sigma ^n$.

Let us now consider the singular points of $\left. \frac d{d\varepsilon
_{j_1}}\cdots \frac d{d\varepsilon _{j_{L+1}}}G_{\vec{\varepsilon}}\right|
_{\varepsilon _1=\cdots =\varepsilon _n=\varepsilon }$. We can simplify
matters by considering the singular points of the related function $%
F_\varepsilon $, which we get by removing some of the analytic factors in
the numerator of (\ref{geuu}): 
\begin{equation}
\sum\limits_{\lambda =-1,+1}\int\limits_{\alpha \in \sigma ^n}\frac{%
d^n\alpha \,\left[ C(\alpha )\right] ^{n-L-1}}{\left[ \sum\limits\Sb %
h=p^2,q^2,q^{\prime 2},  \\ p\cdot q,p\cdot q^{\prime }  \endSb hD_h(\alpha
)+\left( q_0q_0^{\prime }-\lambda \left| \vec{q}\right| \left| \vec{q}^{%
\text{ }\prime }\right| \right) D_{q\cdot q^{\prime }}(\alpha )+i\varepsilon
\cdot C(\alpha )\right] ^{n-L}}.  \label{hred}
\end{equation}
Comparing (\ref{hred}) with the Chisholm result in (\ref{three}), we observe
that $F_\varepsilon $ is the $1+1$ dimensional analog of $g_\varepsilon
(p_0,q,q^{\prime })$.\footnote{%
In $1+1$ dimensions, the exponent of the numerator in (\ref{three}) is $%
n-L-1 $, and the exponent of the denominator is $n-L$.} By this we mean 
\begin{equation}
F_\varepsilon \propto \int d^{2L}l\prod\limits_{j=1}^n\frac{\,1}{k_j^{\sharp
2}+i\varepsilon }+\int d^{2L}l\prod\limits_{j=1}^n\frac{\,1}{k_j^{\flat
2}+i\varepsilon },  \label{jkle}
\end{equation}
where $k_j^{\sharp }$ and $k_j^{\flat }$ are Lorentz vectors in $1+1$
dimensions with the identifications 
\begin{equation}
p^{\sharp \mu }=p^{\flat \mu }=(p_0,0);  \label{liug}
\end{equation}
\begin{equation}
q^{\sharp \mu }=q^{\flat \mu }=(q_0,\left| \vec{q}\right| );  \label{jiuy}
\end{equation}
\begin{equation}
q^{\prime \sharp \mu }=(q_0^{\prime },\left| \vec{q}^{\text{ }\prime
}\right| );\text{ }q^{\prime \flat \mu }=(q_0^{\prime },-\left| \vec{q}^{%
\text{ }\prime }\right| )\text{.}  \label{lkui}
\end{equation}
We conclude that for each monomial $\alpha _{j_1}\alpha _{j_2}\cdots \alpha
_{j_{L+1}}$ appearing in $D_{q\cdot q^{\prime }}(\alpha )$, the singular
points of $\left. \frac d{d\varepsilon _{j_1}}\cdots \frac d{d\varepsilon
_{j_{L+1}}}G_{\vec{\varepsilon}}\right| _{\varepsilon _1=\cdots =\varepsilon
_n=\varepsilon }$ satisfy the $1+1$ dimensional Landau equations for $%
F_\varepsilon $.

Let us now suppose $\left. G_{\vec{\varepsilon}}\right| _{\varepsilon
_1=\cdots =\varepsilon _n=\varepsilon }$ has a singular point that does not
satisfy the Landau equations of $F_\varepsilon $. From our conclusion
regarding the derivatives of $G_{\vec{\varepsilon}}$, we deduce that the
singularities of $\left. G_{\vec{\varepsilon}}\right| _{\varepsilon
_1=\cdots =\varepsilon _n=\varepsilon }$ must have the form 
\begin{equation}
\left. \sum\limits_\eta s_\eta (\varepsilon _{j_1(\eta )},\cdots \varepsilon
_{j_{n(\eta )}(\eta )})\right| _{\varepsilon _1=\cdots =\varepsilon
_n=\varepsilon }  \label{vbwwe}
\end{equation}
where for each $\eta $ and each monomial $\alpha _{\bar{j}_1}\alpha _{\bar{j}%
_2}\cdots \alpha _{\bar{j}_{L+1}}$ appearing in $D_{q\cdot q^{\prime
}}(\alpha ),$ there exists $\bar{j}_i\notin $ $\left\{ j_1(\eta ),\cdots
j_{n(\eta )}(\eta )\right\} $. Let us consider one such term 
\begin{equation}
s_\eta (\varepsilon _{j_1(\eta )},\cdots \varepsilon _{j_{n(\eta )}(\eta )}).
\label{bxc}
\end{equation}
Since $s_\eta $ is independent of some of the $\varepsilon _j$, its
singularities are due to a contracted version of the diagram $\frak{g}$,
where the propagators $\frac{\,1}{k_j^2+i\varepsilon _j}$ corresponding with
internal lines $j\notin \left\{ j_1(\eta ),\cdots j_{n(\eta )}(\eta
)\right\} $ have been removed. We define a new function $\tilde{g}_{\vec{%
\varepsilon}}$ which corresponds with this contracted version of $g_{\vec{%
\varepsilon}}$: 
\begin{equation}
\tilde{g}_{\vec{\varepsilon}}(p_0,q,q^{\prime })=\int
d^{4L}l\prod\limits_{r=1}^{n(\eta )}\frac{\,1}{k_{j_r(\eta )}^2+i\varepsilon
_{j_r(\eta )}}\text{.}  \label{uhbr}
\end{equation}
Performing the loop integrations, we get 
\begin{equation}
\tilde{g}_{\vec{\varepsilon}}(p_0,q,q^{\prime })=\int\limits_{\alpha \in
\sigma ^{n(\eta )}}\frac{d^{n(\eta )}\alpha \,\left[ \tilde{C}(\alpha
)\right] ^{n(\eta )-2L-2}}{\left[ \sum\limits\Sb h=p^2,q^2,q^{\prime 2},  \\ %
p\cdot q,p\cdot q^{\prime },q\cdot q^{\prime }  \endSb h\tilde{D}_h(\alpha
)+i\tilde{C}(\alpha )\sum\limits_{r=1}^{n(\eta )}\alpha _{j_r(\eta
)}\varepsilon _{j_r(\eta )}\right] ^{n(\eta )-2L}},  \label{ewws}
\end{equation}
where 
\begin{equation}
\sigma ^{n(\eta )}=\left\{ \alpha \left| \alpha _j\geq 0,\sum\limits_j\alpha
_j=1,\alpha _j=0\text{ when }j\notin \left\{ j_r(\eta )\right\}
_{r=1}^{n(\eta )}\right. \right\} \text{.}  \label{ytew}
\end{equation}
We define $\tilde{G}_{\vec{\varepsilon}}$ in the same manner as $G_{\vec{%
\varepsilon}}$: 
\begin{equation}
\tilde{G}_{\vec{\varepsilon}}\left( \QTATOP{p_0,q_0,q_0^{\prime }}{\left| 
\vec{q}\right| ,\left| \vec{q}^{\text{ }\prime }\right| }\right) =\frac
1{8\pi ^2}\int d\cos \theta _{\hat{q},\hat{z}}\,d\cos \theta _{\hat{q},\hat{q%
}^{\prime }}\,d\phi _{\hat{q},\hat{z}}\,d\phi _{\hat{q},\hat{q}^{\prime }}\,%
\vec{q}^2\vec{q}^{\text{ }\prime 2}\tilde{g}_{\vec{\varepsilon}%
}(p_0,q,q^{\prime })\text{.}  \label{beew}
\end{equation}
The singular points of $\left. s_\eta \right| _{\varepsilon _1=\cdots
=\varepsilon _n=\varepsilon }$ must satisfy the Landau equations for $\left. 
\tilde{G}_{\vec{\varepsilon}}\right| _{\varepsilon _1=\cdots =\varepsilon
_n=\varepsilon }$. We note, however, that $\tilde{D}_{q\cdot q^{\prime
}}(\alpha )=0$ for all $\alpha \in \sigma ^{n(\eta )}$. $\tilde{g}_{\vec{%
\varepsilon}}(p_0,q,q^{\prime })$ is therefore independent of all the
angular variables and there is a direct relationship between the
singularities of $\left. \tilde{G}_{\vec{\varepsilon}}\right| _{\varepsilon
_1=\cdots =\varepsilon _n=\varepsilon }$ and the singularities of the
corresponding contracted diagram in $1+1$ dimensions: 
\begin{equation}
\tilde{F}_\varepsilon =\int d^{2L}l\prod\limits_{r=1}^{m(\eta )}\frac{\,1}{%
k_{j_r(\eta )}^{\sharp 2}+i\varepsilon }=\int
d^{2L}l\prod\limits_{r=1}^{m(\eta )}\frac{\,1}{k_{j_r(\eta )}^{\flat
2}+i\varepsilon }\text{.}  \label{nree}
\end{equation}
The singular points of $\left. \tilde{G}_{\vec{\varepsilon}}\right|
_{\varepsilon _1=\cdots =\varepsilon _n=\varepsilon },$ and therefore $%
\left. s_\eta \right| _{\varepsilon _1=\cdots =\varepsilon _n=\varepsilon },$
must satisfy the Landau equations for $\tilde{F}_\varepsilon $ (in this
case, no angular integrations are required). But the Landau equations for $%
\tilde{F}_\varepsilon $ are a subset of the Landau equations for $%
F_\varepsilon $, which contradicts our supposition. We conclude that all
singular points of $\left. G_{\vec{\varepsilon}}\right| _{\varepsilon
_1=\cdots =\varepsilon _n=\varepsilon }$ satisfy the Landau equations for $%
F_\varepsilon $.

These results can be generalized to show that for any nonnegative integer $r$%
, the singular points of 
\begin{equation}
\frac 1{8\pi ^2}\int d\cos \theta _{\hat{q},\hat{z}}\,d\cos \theta _{\hat{q},%
\hat{q}^{\prime }}\,d\phi _{\hat{q},\hat{z}}\,d\phi _{\hat{q},\hat{q}%
^{\prime }}\,(\cos \theta _{\hat{q},\hat{q}^{\prime }})^r\vec{q}^2\vec{q}^{%
\text{ }\prime 2}g_\varepsilon (p_0,q,q^{\prime })  \label{yrte}
\end{equation}
must also satisfy the Landau equations for $F_\varepsilon $. When $r>0$, the
argument involving $s_\eta $ is altered slightly$.$ Instead of $s_\eta $
being independent of some of the $\varepsilon _j$'s, $s_\eta $ is now a
polynomial of degree $\leq r$ with respect to these $\varepsilon _j$'s$.$
The arguments that follow, however, are identical with that of the $r=0$
case.

\section{Landau surfaces in $1+1$ dimensions}

We examine the Landau equations for $g_\varepsilon (p_0,q,q^{\prime })$ in $%
1+1$ dimensions, where 
\begin{equation}
g_\varepsilon (p_0,q,q^{\prime })=\int\limits_{\alpha \in \sigma ^n}\frac{%
d^n\alpha \,d^{2L}l\,}{\left[ \sum\limits_{j=1}^n\alpha _jk_j^2+i\varepsilon
\right] ^n}\propto \int d^{2L}l\prod\limits_{j=1}^n\frac{\,1}{%
k_j^2+i\varepsilon }.  \label{ewq}
\end{equation}
Let $\bar{p}^0,\bar{q},\bar{q}^{\prime }$ be a singular point of $%
g_\varepsilon $. There must exist some configuration of the Feynman
parameters $\bar{\alpha}_j\in \sigma ^n$ and internal momenta $\bar{k}_j^\mu 
$ satisfying the Landau equations for $g_\varepsilon $, which we state as
follows.\footnote{%
We are considering a completely massless theory, and the problem of new
singularities at infinite loop momenta (sometimes referred to as second-type
singularities) does not occur.} For each internal line $j$, 
\begin{equation}
\bar{\alpha}_j=0\text{\quad or\quad }\bar{k}_j^2=0  \label{land}
\end{equation}
and for any closed loop $i,$%
\begin{equation}
\sum\limits_jZ_{ij}\bar{\alpha}_j\bar{k}_j^\mu =0\text{,}  \label{dau}
\end{equation}
where $Z_{ij}=\pm 1$ is the relative orientation of $i$ and $\bar{k}_j^\mu $%
. We now fix the values of $\bar{\alpha}_j$ and $\bar{k}_j^\mu $ for the
rest of our discussion. Let us define a new diagram $\frak{g}^{\prime }$,
which we construct by contracting all internal lines $j$ of $\frak{g}$ such
that $\bar{\alpha}_j=0$ or $\bar{k}_j^\mu =0.$ We mention that $\frak{g}%
^{\prime }$ differs from the standard notion for a contracted diagram of $%
\frak{g}$, where one contracts lines satisfying the single condition $\bar{%
\alpha}_j=0$. To avoid confusion, we call $\frak{g}^{\prime }$ the
doubly-contracted diagram of $\frak{g}$. In the following we assume that $%
\frak{g}^{\prime }$ has at least two vertices.

We now add some extra names and markings to the internal lines of $\frak{g}%
^{\prime }$. In $1+1$ dimensions all Lorentz vectors on the light-cone take
the form $(k_0,k_0)$ or $(k_0,-k_0)$ for some real $k_0$. For each internal
line $j$ of $\frak{g}^{\prime }$ we call $j$ a type I line if $\bar{k}_j^\mu
=(\bar{k}_{j0},\bar{k}_{j0})$ and call $j$ a type II line if $\bar{k}_j^\mu
=(\bar{k}_{j0},-\bar{k}_{j0}).$ For each type I line we draw a single
arrowhead, oriented such that the time-component is positive. We do the same
with each type II line, except this time using a double arrowhead. We have
illustrated the use of these markings in Figures 3(a-b). Let $v_1$ and $v_2$
be vertices of $\frak{g}^{\prime }$. If there exists a continuous path along
type I (II) lines from $v_1$ to $v_2$, ignoring for the moment the direction
of the arrowheads, then we say that $v_1$ and $v_2$ are I-connected
(II-connected). If there exists a continuous path along type I (II) lines
from $v_1$ to $v_2$ following along the direction of the arrowheads, then we
say that $v_1$ is I-trajected (II-trajected) to $v_2$. We now prove some
results that follow from the Landau equations.

\begin{proposition}
No vertex is I-trajected (II-trajected$)$ to itself.
\end{proposition}

Proof:\quad Suppose that there existed such an I-trajectory. Let $\bar{k}%
_1^\mu $, $\cdots ,\bar{k}_m^\mu $ be the corresponding momenta of the lines
in the trajectory$.$ By the Landau equations $\sum\limits_{j=1}^m\bar{\alpha}%
_j\bar{k}_j^\mu =0$. This, however, contradicts the fact that each $\bar{%
\alpha}_j$ and $\bar{k}_j^0$ is positive.\quad $\Diamond $

\begin{proposition}
If $v_1$ is I-trajected (II-trajected$)$ to $v_2$, then $v_1$ and $v_2$ are
not II-connected (I-connected$)$.
\end{proposition}

Proof:\quad Suppose $v_1$ is I-trajected to $v_2$, and $v_1$ and $v_2$ are
II-connected$.$ Let us consider the closed loop consisting of an
I-trajectory from $v_1$ to $v_2$ followed by a II-connected path from $v_2$
to $v_1$. We write the momenta of the I-trajectory as $\bar{k}_1^\mu $, $%
\cdots ,\bar{k}_m^\mu $ and the momenta of the II-connected path as $\bar{k}%
_{m+1}^{\prime \mu }$, $\cdots ,\bar{k}_n^{\prime \mu }$. From the Landau
equations, 
\begin{equation}
\sum\limits_{j=1}^m\bar{\alpha}_j\bar{k}_j^\mu =-\sum\limits_{j=m+1}^n\bar{%
\alpha}_j\bar{k}_j^{\prime \mu }.  \label{ntsw}
\end{equation}
The Lorentz vector on the left-side of (\ref{ntsw}) is non-zero since each $%
\bar{\alpha}_j$ and $\bar{k}_j^0$ is positive. But this leads to a
contradiction since the left-side must be parallel to the vector $(1,1)$
while the right-hand side must be parallel to $(1,-1)$.\quad $\Diamond %
\vspace{12pt}$

Since type I vectors and type II vectors are linearly independent, the
constraint of momentum conservation at any internal vertex$,v$, requires
separate conservation of type I vectors and type II vectors. In particular
this means that if there is a type I (II) vector entering $v$ there must
also be a type I (II) vector leaving $v$ and vice-versa. We point out that
these statements are not true of external vertices.

The relation of I-connectivity (II-connectivity) divides the vertices of $%
\frak{g}^{\prime }$ into equivalence classes, which we call I-connected
(II-connected) components. We now prove the following result.

\begin{proposition}
Each I-connected (II-connected) component contains at least two external
vertices.
\end{proposition}

Proof:\quad Let $v$ be an external vertex and let us suppose that $v$ is the
only external vertex in the I-connected component containing $v.$ $v$ must
have a type I vector leaving or entering it. Without loss of generality, we
assume that there is a type I vector leaving $v$ and entering a new vertex
which we call $v_1$. Let us now consider the I-trajectory from $v$ to $v_1$.
Since $v_1$ is an internal vertex, there must also be a type I vector
leaving $v_1$. We can extend the I-trajectory in this manner. By Proposition
1 we never enter the same vertex twice, and since $v$ is the only external
vertex in our I-connected component we can extend the I-trajectory
indefinitely. This contradicts the fact that $\frak{g}^{\prime }$ is finite.

Let us now suppose $v$ is an internal vertex and the I-connected component
containing $v$ has no external vertices. We again consider I-trajectories
starting from $v$ and find the contradiction that the I-trajectory can be
extended indefinitely.$\quad \Diamond \vspace{12pt}$

A simple extension of the argument of the last proof gives us the result
that each I-connected (II-connected) component must contain two external
vertices, $v_1$ and $v_2$ such that $v_1$ is I-trajected (II-trajected$)$ to 
$v_2$. Using this result and Propositions 2 and 3, we find that the
I-connected and II-connected components of $\frak{g}^{\prime }$ must have
the one of the configurations shown in Figures 4(a-h). We have omitted
diagrams related to the ones shown by switching I and II or permuting the
external lines. The topology of these diagrams provides a good deal of
information about the structure of the corresponding Landau surface. One
obvious but useful observation is that if there is a way to cut $\frak{g}%
^{\prime }$ into two pieces such that each cut line is of the same type, the
momentum flowing across the cut must be a Lorentz vector of that type. For
example the Landau surface associated with the diagram in Figure 4(a) is
given by $k^0=k^1\neq 0,$ where $k^\mu $ is the momentum flowing through the
diagram. Examining the structure of each of the diagrams in Figures 4(a-h),
we conclude that the Landau surfaces for any planar $1+1$ dimensional
massless diagram must satisfy one the following properties: (i) one of the
external quark or antiquark lines is light-like; (ii) the total momentum $p$
is light-like (we are using the momentum assignments shown in Figure 1); or
(iii) the total $t$-channel momentum $\frac{q-q^{\prime }}2$ is light-like.%
\footnote{%
In contrast with the previous section, the methods used in this section do
not require that the diagram be planar. If the diagram is not planar, there
is another possibility for the Landau surface, namely that the $u$-channel
momentum $\frac{q+q^{\prime }}2$ is light-like. This can been seen by
twisting the top or bottom half of the diagram shown in Figure 4(h).} This
is the central result of our analysis, and for future reference we will call
these conditions the planar light-cone conditions.

Before closing this section we mention that in addition to the diagrams
shown in Figures 4(a-h), it is also possible that $\frak{g}^{\prime }$
consists of a single point without any internal lines. The Landau surface
associated with this diagram, however, is merely a set of special limit
points on the Landau surfaces associated with the other diagrams. For
example, this includes the point $k^\mu =0$, where $k^\mu $ is the momentum
flowing through the diagram in Figure 4(a)$.$

\section{Non-perturbative OPE}

Our analysis and results can be extended to non-perturbative phenomena by
means of the non-perturbative operator product expansion. In this version of
the operator product expansion one considers, in addition to the usual
perturbative terms, quantities proportional to vacuum condensates of various
local operators.\footnote{%
For a discussion of the non-perturbative operator product expansion see \cite
{nov}.} It is believed that such vacuum condensates are responsible for (or
at least associated with) long-distance confinement phenomena. In this
context the singularities we wish to study are the momentum-space
singularities of the Wilson coefficients in the operator expansion. As an
example we consider the Wilson coefficient of the gluon condensate $%
\left\langle \text{:}G^{\mu \nu }G_{\mu \upsilon }\text{:}\right\rangle $
for the process shown in Figure 5(a). The rectangle shown in Figure 5(a) is
meant to represent the insertion of the condensate $\left\langle \text{:}%
G^{\mu \nu }G_{\mu \upsilon }\text{:}\right\rangle $. The coefficient is
calculated by setting the momentum of each of the two lines touching the
condensate equal to zero. The singularities of this coefficient are the same
as that of the process shown in Figure 5(b), where the $\times $'s are meant
to represent a two point interaction. In fact the singularities of any
connected Wilson coefficient (i.e., one which does not produce
momentum-space delta functions) are equivalent to that of a perturbative
Feynman diagram with some internal lines deleted. Since the methods of our
singularity analysis apply to general planar diagrams, our results apply to
any connected planar Wilson coefficient. We deduce that in $3+1$ dimensions,
the singularities of any angle-integrated, connected, planar Wilson
coefficient satisfy the planar light-cone conditions.

Disconnected Wilson coefficients produce Dirac delta functions in
momentum-space and for this reason we consider them separately. Fortunately
the number of ways the delta function can appear is quite limited, and for
any number of dimensions, the disconnected planar Wilson coefficients are
non-zero only when (i$^{\prime }$) one of the external quark or antiquark
lines has zero momentum; (ii$^{\prime }$) the total momentum $p$ is zero; or
(iii$^{\prime }$) the total $t$-channel momentum $\frac{q-q^{\prime }}2$ is
zero. We note that these conditions for the disconnected coefficients are
subsets of the planar light-cone conditions as defined in section 4. We
conclude that in $3+1$ dimensions, the singularities of any angle-integrated
planar Wilson coefficient, either connected or disconnected, satisfy the
planar light-cone conditions.

\section{Summary and comments}

We found that for massless planar diagrams the singular points of the
angle-integrated amplitude, 
\begin{equation}
\int\limits_{\hat{q},\hat{q}^{\prime }}(\cos \theta _{\hat{q},\hat{q}%
^{\prime }})^r\vec{q}^2\vec{q}^{\text{ }\prime 2}g_{(q,q^{\prime
});\varepsilon }^{ab;cd}(p_0),  \label{uihr}
\end{equation}
lie along the Landau surfaces of the following $1+1$ dimensional analog of $%
g_{(q,q^{\prime });\varepsilon }^{ab;cd}(p_0).$ To each external $3+1$
dimensional vector $e^\mu $ we associate one or two $1+1$ dimensional
vectors $e^{\sharp \mu }$ and $e^{\flat \mu }$, according to the rules 
\begin{equation}
p^{\sharp \mu }=p^{\flat \mu }=(p_0,0),  \label{liug}
\end{equation}
\begin{equation}
q^{\sharp \mu }=q^{\flat \mu }=(q_0,\left| \vec{q}\right| ),  \label{jiuy}
\end{equation}
\begin{equation}
q^{\prime \sharp \mu }=(q_0^{\prime },\left| \vec{q}^{\text{ }\prime
}\right| ),\text{ }q^{\prime \flat \mu }=(q_0^{\prime },-\left| \vec{q}^{%
\text{ }\prime }\right| )\text{.}  \label{lkui}
\end{equation}
The singular points of (\ref{uihr}) are contained in the union of the Landau
surfaces of $g_{(q^{\sharp },q^{\prime \sharp });\varepsilon }^{ab;cd}(p_0)$
and $g_{(q^{\flat },q^{\prime \flat });\varepsilon }^{ab;cd}(p_0)$.

In the $1+1$ dimensional problem, we used the Landau equations to restrict
the topology of the doubly-contracted diagram $\frak{g}^{\prime }$. These
restrictions place constraints on the singular surfaces of the amplitude of
the diagram. We found that the singular surfaces of any planar $1+1$
dimensional massless diagram must satisfy one the following planar
light-cone conditions: (i) one of the external quark or antiquark lines is
light-like; (ii) the total momentum $p$ is light-like; or (iii) the total $t$%
-channel momentum $\frac{q-q^{\prime }}2$ is light-like. Combining this with
the first result, we conclude that the singular surfaces of the
angle-integrated $3+1$ dimensional amplitude must also satisfy the planar
light-cone conditions. We then applied our methods to the non-perturbative
operator product expansion and obtained analogous results for the
singularities of planar Wilson coefficients.

As mentioned in the introduction, confinement is associated with the
infrared behavior of the strong interactions. Knowing the behavior of the
Bethe-Salpeter kernel near its singular surfaces in momentum-space would
provide a great deal of information about the nature of confinement and the
properties of the resulting bound states. Our findings provide a significant
step towards understanding the relation between singularities in the 't
Hooft model and singularities in the $3+1$ dimensional world. Although our
methods apply only to the singularities of finite processes (i.e., finite
loop diagrams and truncated operator product expansions), we point out that
the singularities of the exact Bethe-Salpeter kernel for the massless 't
Hooft model do in fact satisfy the planar light-cone conditions. It is
reasonable to expect that the same might hold true for the exact
angle-integrated Bethe-Salpeter kernel in the $3+1$ dimensional case.
Although the functional form of the singularities of the kernel are not
known, our results indicate the probable location of these singularities and
this information should be quite important in any attempt to build realistic
empirical models of confinement in the relativistic system.

\noindent
{\vspace{12pt}\\ }\Large\bf{Acknowledgements}
\normalsize\rm\\ 

\noindent The author would like to thank Andrew Lesniewski for numerous
discussions on the theory of Landau singularities and Howard Georgi for his
guidance throughout all stages of this project.

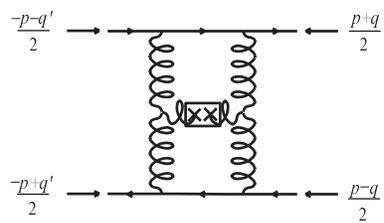
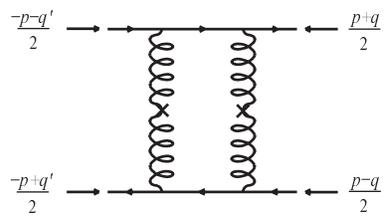
\end{document}